\documentclass[aps,prb,twocolumn]{revtex4}
\usepackage{graphicx}
\usepackage{epsfig}
\begin{document}
\title{Theory of magnetic excitons in the heavy-fermion
superconductor UPd$_{2}$Al$_{3}$}
\author{Jun Chang$^{1}$, I. Eremin$^{1,2}$, P. Thalmeier$^{3}$, and
P. Fulde$^{1}$}
\affiliation{ $^1$ Max-Planck
Institut f\"ur Physik komplexer Systeme, D-01187 Dresden, Germany \\
$^2$ Institute f\"ur Mathematische und Theoretische Physik,
Technische Universit\"at Carolo-Wilhelmina zu Braunschweig, 38106
Braunschweig, Germany \\
$^3$ Max-Planck Institut f\"ur Chemische Physik fester Stoffe,
D-01187 Dresden, Germany }
\date{\today}

\begin{abstract}
We analyze the influence of unconventional superconductivity on the
magnetic excitations in the heavy fermion compound UPd$_2$Al$_3$. We
show that it leads to the formation of a bound state at energies well
below 2$\Delta_0$ at the antiferromagnetic wave vector {\textbf
Q}=$(0,0,\pi/c)$. Its signature is a resonance peak in the
spectrum of magnetic excitations in good agreement with results from
inelastic neutron scattering.  Furthermore we investigate the
influence of antiferromagnetic order on the formation of the
resonance peak. We find that its intensity is enhanced due to
intraband transitions induced by the reconstruction of Fermi surface
sheets.  We determine the dispersion of the resonance peak near {\textbf
Q} and show that it is dominated by the magnetic exciton dispersion
associated with local moments. We demonstrate by a microscopic
calculation that UPd$_2$Al$_3$ is another example in which the
unconventional nature of the superconducting order parameter can be
probed by means of inelastic neutron scattering and determined
unambiguously.
\end{abstract}

\maketitle

\section{Introduction}
The relationship between unconventional superconductivity and magnetism
in heavy-fermion systems and transition metal oxides is one of
the most interesting research areas in condensed matter physics. In
both cases it is widely believed that the magnetic degrees of
freedom play an essential role in the formation of
superconductivity. Furthermore, unconventional superconductivity
yields strong feedback on the magnetic spin excitations in these
systems below the superconducting transition temperature $T_c$. One example is
the famous so-called resonance peak observed in high-$T_c$ cuprates
by means of inelastic neutron scattering (INS) \cite{rossat} whose
nature is still actively debated.\cite{palee} Remarkably, it
has been found that INS reveals the formation of a new magnetic
mode in the superconducting state of the uranium based heavy-fermion
compound UPd$_2$Al$_3$ with $T_c$ = 1.8 K.\cite{sato} Its sharply
peaked intensity, its temperature dependence and the energy position
well below 2$\Delta_0$ (with $\Delta_0$ being the maximum of the
superconducting gap) strongly resembles the resonance peak seen in
high-$T_c$ cuprates. This is particularly remarkable, since the
origin of superconductivity in cuprates and UPd$_2$Al$_3$ seems to
be different. While frequently discussed scenarios in cuprates are a
spin-fluctuation mediated Cooper pairing or the electron-phonon
interaction, in UPd$_2$Al$_3$ a magnetic-exciton mediated pairing
model has been proposed\cite{mchale} based on available
experiments. The latter model is built on the dual nature of the
5$f$ electrons. It consists of localized 5$f^2$ crystalline electric
field (CEF) states which disperse into a magnetic exciton band due
to intersite interactions and a conduction electron band
\cite{zwicknagl} formed by itinerant $5f$ electrons with enhanced
hybridization. The model successfully explains the formation of
unconventional superconductivity in this compound\cite{mchale} based
on the virtual exchange of the magnetic excitons between itinerant
quasiparticles.

It is important to note that the resonant spin excitations in
superconducting cuprates can be seen as a direct consequence of the
$d_{x^2-y^2}$-wave symmetry of the superconducting  order parameter.
Namely, 
the resonance peak occurs only if the order parameter changes sign
in the first Brillouin zone (BZ).\cite{respeak} Thus, INS can be
considered as a bulk probe for the unconventional nature of
superconductivity in these compounds. Therefore it is important to
search for such an effect in other unconventional superconductors as
well. In this paper we analyze the consequences of the
unconventional pairing on the magnetic excitations in UPd$_2$Al$_3$.
We show that in addition to the magnetic exciton dispersion present
in the normal state, unconventional superconductivity induces the formation of a bound state below
$T_c$ with an associated resonance peak in the magnetic spectrum at the antiferromagnetic (AF) wave
vector {\textbf Q}=$(0,0,\pi/c)$ where $c$ is a lattice constant
along the crystallographic $z$ axis. Its frequency is well
below 2$\Delta_0$ and in good agreement with experimental data. We
show that similar to cuprates the resonance peak in
UPd$_2$Al$_3$ is a consequence of an unconventional superconducting
order parameter which changes sign at regions of the Fermi surface
connected by the antiferromagnetic wave vector {\textbf Q}. We analyze
the influence of antiferromagnetism on the formation of the
resonance peak and surprisingly find that its intensity is enhanced
due to the reconstruction of the Fermi surface. We find that the
dispersion of the resonance peak away from {\textbf Q} is controlled by
the momentum dependence of excitations of the localized magnetic
moment (magnetic exciton).

The resonance peak in UPd$_2$Al$_3$ has also been studied by Bernhoeft {\textit et al.}\cite{bernhoft1} within a phenomenological two
component spin susceptibility model. However, in our microscopic
calculations we show that antiferromagnetic order plays a crucial
role in the formation of the resonance peak below $T_c$.

\section{The Hamiltonian}

Following previous consideration by McHale {\textit et al.}
\cite{mchale} we use the low-energy Hamiltonian describing the
interaction of the itinerant $f$ electrons and magnetic excitons
originating from localized 5$f^2$ crystalline electric field (CEF)
states:
\begin{eqnarray}
H_0 & = &{\sum_{{\mathbf p}\sigma}}\xi_{\mathbf p}c_{{\mathbf
p}\sigma}^{\dagger}c_{{\mathbf p}\sigma}+ {\sum_{\mathbf q}}\omega_{\mathbf
q}\left(\alpha_{\mathbf q}^{\dagger}\alpha_{\mathbf q}+ \frac{1}{2}\right)
\nonumber \\
&-& \frac{g}{N}{\sum_{\mathbf {p,q}}}c_{{\mathbf
p}\alpha}^{\dagger}\mathbf{\sigma}_ {\alpha\beta}^{z}c_{{\mathbf
{p+q}},\beta}\lambda_{\mathbf q}\left(\alpha_{\mathbf q}+\alpha_{-{\mathbf
q}}^{\dagger}\right), \label{eq:hamilt}
\end{eqnarray}
where $\lambda_{\mathbf q}^2=\frac{\Delta_{\rm CEF}}{\omega_{\mathbf q}}$, and
$\Delta_{\rm CEF}=6$ meV is the energy gap of the 5$f^2$ electrons
between the ground and first excited states in the crystalline
electric field. The dispersion of the magnetic excitons is
approximately described by $\omega(q_z)=\omega_{ex}[1+\beta\,\cos
(cq_z)]$ with $0<\beta \simeq1$, where $g$ is the coupling constant
between the itinerant electrons and the localized magnetic moments.
We adopt parameter values $\omega_{ex}$ = 5.5 meV and $\beta=0.72$.
Note, here we follow Ref. \onlinecite{mchale} in assuming that only
the $\sigma_z$ component of the conduction electrons can excite
magnetic excitons. Therefore the spin-space isotropy is broken in a
maximal (Ising) way. As a result the usual classification of
Cooper pairs into spin-triplet and spin-singlet states is not valid
and the notation { \textit  equal and opposite spin pairing} states
should be better used instead. However, we will still speak of
singlet and triplet Cooper-pairing states as commonly done.

Eq.\ (\ref{eq:hamilt}) gives rise to  fermionic and bosonic
self-energies and is particularly relevant for electron-hole states
 separated by the antiferromagnetic wave vector {\textbf
Q}=(0,0,$\pi/c$). Previously it has been shown that this interaction
explains superconductivity in UPd$_2$Al$_3$.\cite{mchale} We define
the electron and magnetic exciton Green's functions as follows:
\begin{eqnarray}
G_{\sigma \, \sigma'} ({\mathbf p}, i \omega_m )& = & - \left\langle
T_{\tau}c_{\mathbf p \sigma }(\tau) c_{\mathbf p
\sigma'}^{\dagger}(0)\right\rangle_{\rm FT},
 \nonumber \\
D\left({\mathbf q},iv_{n}\right )& = & - \left\langle T_{\tau} a_{\mathbf
q}(\tau)a_{\mathbf {-q}}^{\dagger}(0)\right\rangle_{\rm FT},
\end{eqnarray}
where $a_{\mathbf q} (\tau) = \alpha_{\mathbf q}(\tau) +
\alpha^{\dagger}_{\mathbf -q}(\tau)$ and $D$ is essentially the
pseudospin susceptibility. The bare magnetic exciton Green's
function is given by $ D_0\left({\mathbf q},i\nu_{n}\right ) = -
\frac{\Delta_{\rm CEF}}{2} \frac{1}{\nu_n^2 + \omega^2_{\mathbf q}}$. Due to
the interaction of the magnetic excitons with conducting electrons,
the feedback effect on the former results in
\begin{eqnarray}
D=\frac{D_{0}}{1-D_{0}\Pi_{0}}=-\frac{2\omega_{\mathbf
q}}{\omega^{2}-\omega_{\mathbf q}^{2}+2\omega_{\mathbf q}\Pi_{0}} \quad,
\label{eq:exciton}
\end{eqnarray}
where the magnetic exciton self-energy is given by
\begin{eqnarray}
\Pi_{0}({\mathbf q},i\omega_{n})=g^{2}\frac{\Delta_{\rm CEF}}{\omega_{\mathbf
q}}\chi_{0}({\mathbf q},i\omega_{n}) \quad.
\end{eqnarray}
Here, the spin susceptibility of the conduction electrons in the
superconducting state is
\begin{widetext}
\begin{eqnarray}
\chi_{0}({\mathbf q},i\omega_{n})  = -\frac{1}{\beta}\sum_{i\omega_m,
{\mathbf k}}\left[G({\mathbf k},i\omega_{m})G^{\dagger}({\mathbf
{k+q}},i\omega_{m}+i\omega_{n})+ F({\mathbf
k},i\omega_{m})F^{\dagger}({\mathbf {k+q}},i\omega_{m}+i\omega_{n})\right],
\label{eq.sus0}
\end{eqnarray}
where bare Green's functions of superconducting electrons are
$G({\mathbf k},i\omega_{m})=-\frac{i\omega_{m}+\xi_{{\mathbf k}}}{\omega_{m}^{2}+\xi_{{\mathbf k}}^{2}+\Delta_{{\mathbf k}}^{2}}$,
$F({\mathbf k},i\omega_{m})=\frac{\Delta_{{\mathbf k}}}{\omega_{m}^{2}+\xi_{{\mathbf k}}^{2}+\Delta_{{\mathbf k}}^{2}}$.
A straightforward evaluation of the sum over the Matsubara
frequencies gives (at $T=0$ K)
\begin{eqnarray}
\mbox{Im}\,\chi_{0}({\mathbf
q},\omega)=\frac{1}{4}\frac{1}{(2\pi)^{3}}\int d^3 k
\left(1-\frac{\xi_{{\mathbf {k+q}}}\xi_{{\mathbf k}}+\Delta_{{\mathbf k}}\Delta_{{\mathbf {k+q}}}}
{E_{{\mathbf {k+q}}}E_{{\mathbf k}}}\right)\delta\left(\omega-E_{{\mathbf {k+q}}}-E_{{\mathbf k}}\right) \quad.
\label{susT0}
\end{eqnarray}
\end{widetext}
The Fermi surface of the conducting electrons is almost like a
cylinder with weak dispersion along the $z$ direction. Neglecting
the anisotropy of dispersion in the plane yields $ \xi_{\mathbf
k}=\epsilon_{k_{\bot}}+\epsilon_{k_{z}}-\mu=\epsilon_{\bot}w^{2}+\epsilon_{\parallel}\,\cos(ck_{z})-\mu
$ ($\epsilon_{\parallel}\ll \epsilon_{\bot}$, $w=k_{\bot}/k_0 \leq 1
$) and $E_{\mathbf k}=\sqrt{\ \xi_{\mathbf k}^{2}+\Delta_{\mathbf k}^{2}}$.
Here, we approximate the hexagonal unit cell by a circle with radius
$k_0$ chosen so that the hexagon and the circle have the same area.
Furthermore, we assume a parabolic dispersion in the plane. Due to
the Ising-type anisotropy of the interaction between conduction
electrons and magnetic excitons it has been previously
found\cite{mchale} that both pure paramagnetic, {\textit i.e.},
spin-singlet states $\frac{1}{\sqrt{2}}\left( | \uparrow \downarrow
\rangle - | \downarrow \uparrow \rangle \right)$ with $\cos (ck_z)$
momentum dependence and spin-triplet states ($S_{z}=0$) $
\frac{1}{\sqrt{2}}\left( | \uparrow \downarrow \rangle + |
\downarrow \uparrow \rangle \right)$ with $\sin (ck_z) $ have the
highest (degenerate) superconducting transition temperature. Thus,
in the following we will consider the two superconducting order
parameters, $\Delta^s_{\mathbf k}=\Delta_0\,\cos (ck_z)$ and
$\Delta^{t}_{\mathbf k}=\Delta_0\,\sin (ck_z)$ as the most relevant ones
in this model.

Let us now discuss the consequences of the behavior of Im$\,\chi_0$
for the magnetic exciton dispersion which follows from Eq.\ 
(\ref{eq:exciton}). The dispersion of the magnetic exciton in the
presence of coupling to the conduction electrons is given by
\begin{equation}
\omega^2 = \omega_{\mathbf q}^2 - 2 g^2 \Delta_{\rm CEF} \mbox{Re}\,\chi_0
({\mathbf q}, \omega)  \quad. \label{eq:pole}
\end{equation}

In the normal paramagnetic state, Re$\,\chi_0$ is a constant at low
frequencies determined in our case by the curvature of the Fermi
surface along the $k_z$ direction. At the same time Im$\,\chi_0 \propto \
-i \gamma \omega$, where $\gamma$ is a Landau damping constant. Thus,
the bare magnetic exciton acquires a linewidth and renormalizes
slightly by a certain constant which changes the original position
of $\omega_{\mathbf q}$ downwards. In the superconducting state the
renormalization is strongly dependent on the gap symmetry. For
example, in the conventional $s$-wave state Re$\,\chi_0$ is less
than its normal state value. First, the $s$-wave superconducting
gap gives a negative contribution to $\chi_0$ as follows from the
coherence factor in Eq.\ (\ref{susT0}), and second it yields the
spin excitation gap structure in Im$\,\chi_0$ as follows from the
$\delta$ function. Therefore, the feedback of the conduction
electrons on the magnetic exciton becomes weaker and yields a shift
of the magnetic exciton dispersion towards higher frequencies in the
superconducting state with respect to its normal state value.

\begin{figure}[h]
\epsfig{file=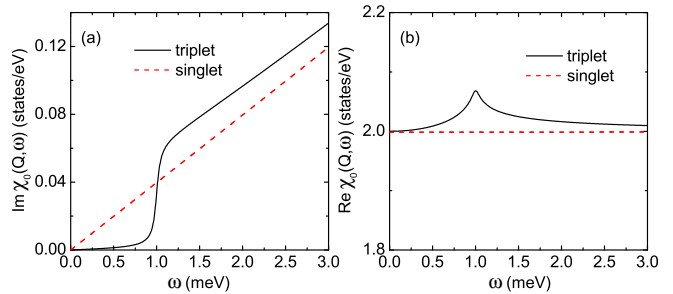,width=1.0\columnwidth,angle=0}
\caption{(Color online) The (a) imaginary and (b) real parts of the
longitudinal component of the conducting electrons spin
susceptibility $\chi_0({\mathbf Q},\omega)$  as a function of frequency
for  singlet (red dashed curve) and triplet (black solid curve)
Cooper pairing. The curves for the normal state (not shown) almost
coincide with those for the spin singlet superconducting state. Here
and in the following we use $T=0.15$ K, $\Delta_0 = 0.5$ meV,
$\epsilon_{\parallel} = 25$ meV and a damping parameter 35 $\mu$eV
for the numerical calculations.} \label{fig1}
\end{figure}

At the same time, the unconventional character of the
superconducting gap entails immediate consequences for Eqs.\ 
(\ref{susT0}) and (\ref{eq:pole}). Namely, at the Fermi surface
(i.e., for $\xi_{\mathbf k} = 0$) one finds $\Delta^s_{\mathbf
{k+Q}}=-\Delta^s_{\mathbf k}$ as well as $\Delta^{tr}_{\mathbf
{k+Q}}=-\Delta^{tr}_{\mathbf k}$ where {\textbf Q} is the antiferromagnetic
wave vector. Thus, in both cases the anomalous coherence factor
equals $2$ for all $k_z$ momenta which is in strong contrast to the
usual $s$-wave symmetry of the superconducting gap. Simultaneously,
the $\delta$ function in Eq.\ (\ref{susT0}) starts to contribute for
$\Omega_{cr} (k_z, q_z) =\text{Min}\left(E_{{\mathbf k}+\hat{\mathbf z} q_z}+E_{\mathbf k}\right)=\sqrt{\epsilon^2_{\parallel}
\left[\cos(ck_{z}+cq_{z})-\cos(ck_{z})\right]^2
 + (|\Delta_{k_z}|+|\Delta_{ k_z+q_z}|)^2}$ which is determined by the value of the
 gap, $\Delta_{\mathbf k}$
 and the quasiparticle dispersion along $k_z$ axis, $\epsilon_{||}$.
Note that $\Omega_{cr}^s (k_z,q_z=\pi/c) =
 2| \cos (ck_z) |\sqrt{\Delta_0^2
+\epsilon_{\parallel}^2} $
and $\Omega_{cr}^{tr} (k_z,q_z=\pi/c) =
 2\sqrt{\Delta_0^2+\left(\epsilon_{\parallel}^2
-\Delta_0^2\right) \cos^2 (ck_z) }$ for the singlet and triplet
superconducting gap, respectively. It is interesting to note that
for the singlet case there is no gap in Im$\,\chi_0$ and therefore, it
is nearly the same as in the normal state\cite{respeak} [see
Fig.\ \ref{fig1}(a)]. Then Re$\,\chi_0$ is a constant at low
frequencies. Therefore the magnetic exciton dispersion will be
simply shifted in proportion to the total value of Re$\,\chi_0$
exactly as in the normal state. Correspondingly, no
strong feedback on the magnetic exciton due to superconductivity
takes place.

However, one sees that in the case of the triplet order parameter Im
$\,\chi_0$ will be gapped at least up to values of
%
%
$\Omega_{cr}^{tr}=|2\Delta_0|$($\epsilon_{\parallel}\gg \Delta_0$). Then due to a combined effect of the
anomalous coherence factor and the $\delta$ function, a
discontinuous jump in Im$\,\chi_0$  occurs at about $\omega = 2
\Delta_0$ for the triplet order parameter. Via Kramers-Kronig
transformation the discontinuous jump in Im$\,\chi_0$ yields  a
logarithmic singularity in Re$\,\chi_0$. Note, the logarithmic singularity in Fig.\ 1.(b) is suppressed by a weak damping. Furthermore, below
$\Omega_{cr}^{tr}$ the Re$\,\chi_0$ is increased with respect to its
normal state value for $\omega \neq 0$ as shown in
Fig.\ \ref{fig1}(b).

A frequency dependence of the Re$\,\chi_0({\mathbf Q}, \omega)$ can yield
more than one solution of Eq.\ (7). In order to demonstrate how those
solutions can be found above and below T$_c$ we illustrate in Fig.\ 2
the possible characteristic behavior of $\omega_{\mathbf
q}^2-2g^2\Delta_{\rm CEF}\,\mbox{Re}\,\chi_0(\mathbf q, \omega)$.
%
\begin{figure}[h]
\centerline{\epsfig{file=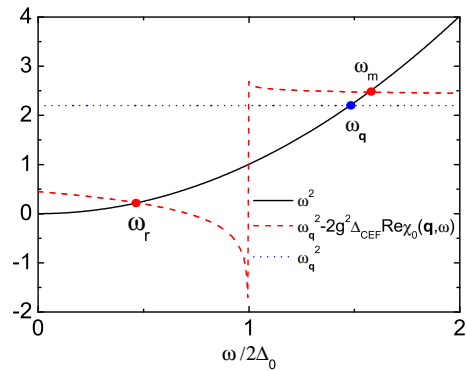,width=0.7\columnwidth,angle=0}}
\caption{(Color online) Illustration of the solutions of
Eq.\ (\ref{eq:pole}) at wave vector {\textbf q = Q}. In the normal
state there is only one crossing point between the $\omega^2$ curve
(solid) with the $\omega_{\mathbf q}$ line (dotted) yielding the
frequency of the magnetic exciton. Note that $\omega_{\mathbf q}$ may
slightly differ from the bare exciton dispersion due to Re$\,\chi_0
({\mathbf Q}, \omega)$ = const in the normal state. In the
superconducting state due to the strong frequency dependence of
Re$\,\chi_0 ({\mathbf q}, \omega)$ (and/or Re$\,\Pi_0$) one finds several
intersecting points of $\omega^2$ with $\omega_{\mathbf
q}^2-2g^2\Delta_{\rm CEF}\mbox{Re}\,\chi_0(\mathbf q, \omega)$. The lowest
pole ($\omega_r < 2 \Delta_0$) occurs at very small damping
(Im$\,\chi_0$ is zero or small) resulting in a resonancelike peak in
Im$\,D({\mathbf q}, \omega)$. The second crossing point is not visible in
Im$\,D$ due to a large peak in Im$\,\chi_0$ or strong damping around
$2\Delta_0$. The third crossing point, $\omega_m$ occurs at energies
larger than $2\Delta_0$ and represents the feedback effect of
superconductivity on the magnetic exciton.} \label{fig2}
\end{figure}
Here, we assume that in normal state the Re$\,\chi_0$ is almost frequency independent  and magnetic exciton's peak position shifts slightly in superconducting state with respect to its normal state value. Due to the gap structure in superconducting state, depending on the value of $g$, a new pole may occur at energies less than $2\Delta_0$. If Im$\,\chi_0$ is small or zero at these frequencies, the total
Im$\,D$, i.e., the spectral function of magnetic excitations, shows a
resonance peak which occurs only in the superconducting state.  This agrees well with experimental INS data\cite{sato}. Moreover, at
higher energies one finds in addition two more poles in
Eq.\ (\ref{eq:pole}). The latter yields an additional structure in
Im$\,D$ which is a renormalized magnetic exciton with finite damping. This typical behavior of the susceptibility  can be found in Fig.\ 3.

So far we have ignored the coexistence of antiferromagnetism and
superconductivity in UPd$_2$Al$_3$ for the conduction
electrons.\cite{krimmel} Antiferromagnetic order results in
UPd$_2$Al$_3$ due to the interaction of neighboring uranium ions.
This leads to a dispersion for exciton state and eventually to an
antiferromagnetic instability and a new ground state.\cite{thalmeier} The unit cell
is doubled and the Brillouine zone is correspondingly reduced. The
new dispersion of the conducting electrons enters the expression for
the spin susceptibility. Then the solutions of Eq.\ (7) must be
redetermined. This is done in the following.

The total Hamiltonian is
\begin{eqnarray}
H=H_0 + m{\displaystyle \sum_{\mathbf p}}\sigma c_{{\mathbf
{p+Q}}\sigma}^{\dagger}c_{{\mathbf p}\sigma} \label{eq:totalH}
\end{eqnarray}
where $m$ denotes the value of the effective antiferromagnetic
staggered field. This term leads to a splitting of the quasiparticle
energy dispersion into two bands.\cite{amici} In particular, the
Hamiltonian (\ref{eq:totalH}) can be easily diagonalized by a
unitary transformation\cite{ismer} and the resulting energy
dispersions are $E_{\mathbf k}^{\pm}=\sqrt{(\varepsilon_{\mathbf
k}^{\pm})^{2}+\Delta_{\mathbf k}^{2}}$ with $\varepsilon_{\mathbf k}^{\pm}=
[\epsilon^+_{\mathbf k}\pm \sqrt{(\epsilon_{\mathbf
k}^{-})^2+m^{2}}]$. Here, we have introduced $\epsilon_{\mathbf
k}^{+}=\frac{1}{2}\left(\xi_{\mathbf k}+\xi_{\mathbf {k+Q}}\right)$ and
$\epsilon_{\mathbf k}^{-}=\frac{1}{2}\left(\xi_{\mathbf k}-\xi_{\mathbf
{k+Q}}\right)$. The AF Fermi surface of the two bands
$\varepsilon_{\mathbf k}^{\pm}$ consists of two disjoint cylinders in
the reduced AF BZ $|p_{z}|\leq \frac{\pi}{c}$ (see Ref.
\onlinecite{cylinders}). In a pure AF state, the real part of the
susceptibility $\chi_0$ at low energies is determined by the
intraband processes and can be approximated by $\chi_0 ({\mathbf q},
\omega) \propto \sum_{\mathbf k} A_{\mathbf k,q}
\frac{f(\varepsilon^{\pm}_{\mathbf k}) - f(\varepsilon^{\pm}_{\mathbf
{k+q}})}{\varepsilon^{\pm}_{\mathbf {k+q}} - \varepsilon^{\pm}_{\mathbf k} +
\omega}$ where $A_{\mathbf k,q}$ is the AF coherence factor. At wave
vector {\textbf Q} for $\omega = 0$ the susceptibility is proportional
to the density of states and decreases rapidly to zero as one
increases frequency. This is a consequence of the equality
$\varepsilon^{\pm}_{\mathbf k} = \varepsilon^{\pm}_{\mathbf {k+Q}}$. Therefore
the renormalization of the magnetic excitons due to conduction
electrons can be safely ignored.

\begin{figure}[h]
\centerline{\epsfig{file=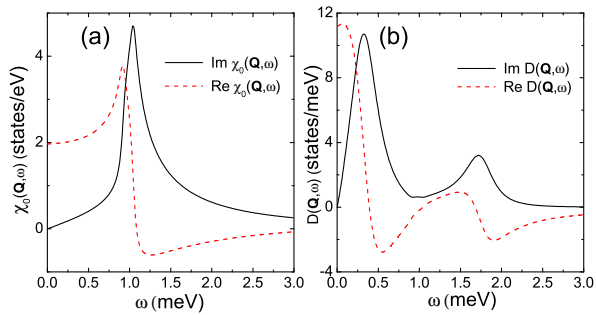,width=0.9\columnwidth,angle=0}}
\caption{(Color online) Results for $\chi_0({\mathbf Q},\omega)$ and
$D({\mathbf Q}, \omega)$ in the coexisting AF and singlet
superconducting state. (a) Calculated real and imaginary parts of
the longitudinal component of the conduction electron spin
susceptibility, $\chi_0({\mathbf Q},\omega)$. (b) The real and imaginary
part of the total pseudospin susceptibility $D({\mathbf Q}, \omega)$.
Here $m = 50$ meV, $g = 10$ meV. We use the gap
function $\Delta_{\mathbf k} \propto \Delta_0 \left[\cos (ck_z) -\frac 15
\cos (3ck_z) +\frac{1}{30} \cos (5 ck_z) \right]$( Ref.\ 
\onlinecite{bernhoft1} ). } \label{fig3}
\end{figure}

Most importantly, in the superconducting state coexisting with AF,
the imaginary part of the spin susceptibility of the conduction
electrons including intraband and interband scattering is given for
$T$=0 K, ${\mathbf q = Q}$, and $\omega>0$ by
\begin{widetext}
\begin{eqnarray}
\lefteqn{\mbox{Im}\,\chi_0({\mathbf Q}, \omega)  =} && \nonumber
\\ && \sum_{\mathbf k}\frac{1}{8} \, \delta\left(\omega-2E_{\mathbf k}^{+}\right)
\left(1-\frac{\left(\varepsilon_{\mathbf k}^{+}\right)^{2}-\Delta_{\mathbf
k}^{2}}{\left(E_{\mathbf
k}^{+}\right)^{2}}\right)\left(\frac{2m^{2}}{\epsilon_{k_{z}}^{2}+m^{2}}\right)
 +\sum_{\mathbf k}\frac{1}{4}\, \delta\left(\omega-E_{\mathbf k}^{+}-E_{\mathbf
k}^{-}\right)\left(1-\frac{\varepsilon_{\mathbf k}^{+}\varepsilon_{\mathbf
k}^{-}-\Delta_{\mathbf k}^{2}}{E_{\mathbf k}^{+}E_{\mathbf
k}^{-}}\right)\left(\frac{2\epsilon_{k_{z}}^{2}}{\epsilon_{k_{z}}^{2}+m^{2}}\right)\nonumber
\\
&&+\sum_{\mathbf k}\frac{1}{8}\, \delta\left(\omega-2E_{\mathbf
k}^{-}\right)\left(1-\frac{\left(\varepsilon_{\mathbf
k}^{-}\right)^{2}-\Delta_{\mathbf k}^{2}}{\left(E_{\mathbf
k}^{-}\right)^{2}}\right)\left(\frac{2m^{2}}{\epsilon_{k_{z}}^{2}+m^{2}}\right)
\quad. \label{eq:susAF}
\end{eqnarray}
\end{widetext}
Here the first and the third terms describe the intraband
quasiparticle pair creation while the second term refers to the
corresponding interband process.  Note that Eq.\ (\ref{eq:susAF})
contains two types of coherence factors, {\textit i.e.}, due to
superconductivity and antiferromagnetic order, respectively. As
usual, the low-energy behavior of Im$\,\chi_0$ is dominated by the
intraband contributions. We also assume that the presence of
antiferromagnetism does not change the qualitative behavior of the
superconducting gap, {\textit i.e.}, the position of the line node and
the corresponding change of sign of the superconducting order
parameter remain the same although some higher harmonics may
appear.\cite{amici} As in Eq.\ (\ref{susT0}), the
superconducting coherence factors equal 2 for $k_z$ momenta close to
the Fermi surface. At the same time, the reconstructed conduction
bands in the AF state have only one-half the original period, i.e.,
$\varepsilon_{\mathbf k}^{\pm} = \varepsilon_{\mathbf {k+Q}}^{\pm}$. Therefore
all parts of the Fermi surface can  be connected by the
antiferromagnetic wave vector {\textbf Q} and simultaneously have a sign
change $\Delta_{\mathbf {k+Q}} = -\Delta_{\mathbf k}$ of the gap except at the
nodal points. As a result, Im$\,\chi_0({\mathbf Q},\omega)$ is non-zero
for $\omega > 0$ due to the contribution of the nodal states both in
the singlet and the triplet Cooper-pairing cases. With increasing
frequency Im$\,\chi_0$ increases up to energies of about $2\Delta_0$
and then decreases [see Fig.\ \ref{fig3}(a)]. The functional
dependence of Im$\,\chi_0$ at low frequencies resembles the behavior
of the density of states except that the structure occurs at around
$2\Delta_0$.
%
%
Correspondingly, the real part of $\chi_0 ({\mathbf Q},\omega = 0)$ is
the same as in a pure AF state. However, away from $\omega = 0$ it
does not drop as in the pure AF state but increases quadratically up to
about 2$\Delta_0$ due to the structure of Im$\,\chi_0$ induced by the
superconducting gap. Only then does Re$\,\chi_0$ drop to small
values. Altogether Re$\,\chi_0 ({\mathbf Q}, \omega)$ increases in the
superconducting state for $\omega < 2\Delta_0$ due to the
unconventional nature of the superconducting order parameter.
However, the pure resonance (bound state) in Im$\,D$ is not realized
due to finite damping. An additional pole in Im$\,D$ still exists
at frequencies smaller than 2$\Delta_0$ due to a strong increase of
Re$\,\chi_0$ at small frequencies as shown in Fig.\ \ref{fig3}(b). At
higher frequencies it becomes small and thus
 another pole appears corresponding to the broadened original magnetic
 exciton. Thus, Im$\,D$ has a two-pole structure as shown in
 Fig.\ \ref{fig3}(b). Note, in order to increase the intensity of the low-energy pole in
 Im$\,D$ we include the contribution of the
 higher harmonics to the gap function. As already mentioned, they
 are due to the presence of AF order \cite{amici}.

\begin{figure}[h]
\centerline{\epsfig{file=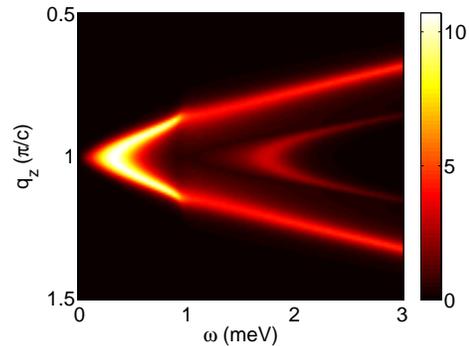,width=0.7\columnwidth,angle=0}}
\caption{(Color online) Contour plot of the imaginary part of the
total pseudospin susceptibility as function of frequency, $\omega$
and $q_z$ momentum. One clearly observes two distinct peaks at {\textbf
Q}. The one at low energies represents the resonance peak induced by
the feedback of superconductivity and the one at higher $\omega$ is
the renormalized magnetic exciton. Away from {\textbf Q} both peaks
disperse upward in energy following the behavior of the normal state
magnetic exciton.} \label{fig4}
\end{figure}

 Finally we discuss the dispersion of the magnetic excitations away
 from {\textbf Q} along the $q_z$ direction. In Fig.\ \ref{fig4} we show the calculated momentum and
frequency dependencies of Im$\,D(q_z, \omega)$.
 It is clear that as soon as $q_z \neq {\pi /c}$ the
%
%
original magnetic exciton has a strong upward dispersion in the
normal state. Therefore, effects connected with renormalization
induced by superconductivity will also be shifted towards higher
energies. The pole induced by superconductivity shows
dispersion similar to the magnetic exciton [see Fig.\ \ref{fig4}].
For both singlet and triplet order parameters our results are in
fair agreement with recent INS data.\cite{heiss} Namely, in the
superconducting state one finds two distinct energy dispersions, one
being the resonancelike feature with high intensity inside the
superconducing gap and the second, that of localized magnetic
excitons renormalized by the conducting electrons. Another
interesting point worth noting is that due to the doubling of the
unit cell and the equality $\varepsilon^{\pm}_{\mathbf k} =
\varepsilon^{\pm}_{\mathbf {k+Q}} $, the effect of the $\sin (ck_z)$ and
$\cos (ck_z)$ gaps leads to a very similar behavior for Im$\,\chi_0$. The
slight difference in the absolute magnitude arises from the
different densities of states at those regions of the Fermi surface
where the maximum of the singlet and the triplet gaps occur.
Altogether this does not change the functional form of Im$\,\chi_0$.
We mention that thermal conductivity results in a rotating field
\cite{watanabe} are compatible with both $\cos (ck_z)$ and $\sin (ck_z)$
order parameters while
 the observed Knight shift \cite{kitaoka} seems to favor the former.
Interestingly, we also found that a recently proposed
superconducting gap with $\cos (2 ck_z)$ symmetry \cite{maki} does not
lead to the formation of low-energy spin excitations around wave
vector {\textbf Q} in the superconducting gap. The reason is that its
momentum dependence yields no change of the sign of the
superconducting order parameter, $\Delta_{\mathbf k} =\Delta_{\mathbf {k+Q}}$,
and thus no constructive contribution can result from the anomalous
coherence factor.

In conclusion, we have investigated the effects of superconductivity
on the magnetic excitations in the unconventional superconductor
UPd$_2$Al$_3$. In particular, due to the change in sign of the
superconducting order parameter the conduction electron
susceptibility is enhanced in the superconducting state which yields
an additional pole (bound state) in the total susceptibility. We
further analyzed the role played by  antiferromagnetism and found
that its presence increases the spectral weight of the resonance due
to the doubling of the unit cell. However, the resonance peak in the
AF phase becomes a virtual bound state due to finite damping.
Finally we point out that UPd$_2$Al$_3$ is another known example
where the unconventional nature of the superconducting order
parameter yields a structure in the magnetic susceptibility as in
layered high-$T_c$ cuprates. Therefore it can be regarded as a model
system of unconventional superconductivity studied by inelastic
neutron scattering.

The authors acknowledge helpful discussions with G. Zwicknagl, J.-P. Ismer,
and A. Klopper.

\end{document}